\documentclass[11pt]{article}
\usepackage{amssymb}
\usepackage{epsfig}

\newcommand{\BE}{\begin{equation}}
\newcommand{\EE}{\end{equation}}
\newcommand{\BA}{\begin{eqnarray}}
\newcommand{\EA}{\end{eqnarray}}

\begin{document}
\begin{titlepage}

\vspace*{1mm}
\begin{center}

            {\LARGE{\bf The motion of the Solar System and \\
the Michelson-Morley experiment }}

\vspace*{14mm}
{\Large  M. Consoli and E. Costanzo}
\vspace*{4mm}\\
{\large
Istituto Nazionale di Fisica Nucleare, Sezione di Catania \\
Dipartimento di Fisica e Astronomia dell' Universit\`a di Catania \\
Via Santa Sofia 64, 95123 Catania, Italy}
\end{center}
\begin{center}
{\bf Abstract}
\end{center}
Historically, 
the Michelson-Morley experiment has played a crucial role for abandoning
the idea of a preferred reference frame, the ether, and for replacing
Lorentzian Relativity with Einstein's Special Relativity. However, 
our re-analysis of the Michelson-Morley
original data, consistently with the point of view already
expressed by other authors, shows that the experimental observations
have been misinterpreted. Namely, the fringe shifts 
point to a non-zero {\it observable} Earth's velocity
$ v_{\rm obs} = 8.4 \pm 0.5~ {\rm km/s} $. 
Assuming the existence of a preferred reference
frame, and using Lorentz transformations to extract the kinematical 
Earth's velocity that corresponds to this $v_{\rm obs}$, we obtain a
{\it real} velocity, in the plane of the interferometer,
$v_{\rm earth} = 201 \pm 12~{\rm  km/s}$. This value 
is in excellent agreement with Miller's calculated value
$v_{\rm earth} = 203 \pm 8$~{\rm  km/s} and suggests
that the magnitude of the 
fringe shifts is determined by the typical velocity of the
Solar System within our galaxy. This conclusion, which is also 
consistent with the
results of {\it all} other classical experiments, leads to
an alternative interpretation of the Michelson-Morley type of experiments. 
Contrary to the generally accepted ideas of last
century, they provide experimental evidence for the existence of a 
preferred reference frame. This point of view is also consistent 
with the most recent data for the anisotropy of the 
two-way speed of light in the vacuum.  
\vskip 35 pt
\end{titlepage}

\section{Introduction}

The Michelson-Morley experiment \cite{mm} was designed
to detect the relative motion of the Earth with respect to a preferred 
reference frame, 
the ether, by measuring the shifts of the fringes in an optical interferometer.
These shifts, that should have been proportional to the square of the 
Earth's velocity, were found to be much smaller than expected. Thus,
that experiment was taken as an evidence that there is no ether and, as such,
represented the essential ingredient 
for deciding between Lorentzian Relativity and
Einstein's Special Relativity. 

However, according to some authors, 
the fringe shifts observed by Michelson and Morley, while
certainly smaller than the classical prediction corresponding to the orbital
velocity of the Earth, were {\it not} negligibly small. 
This point was clearly expressed by Hicks, see page 36 of Ref.\cite{hicks}
"..the numerical data published in the Michelson-Morley 
paper, instead of giving a null
result, show a distinct evidence of an effect of the kind to be expected" 
and also by Miller, see Fig.4 of Ref.\cite{miller}. In the latter case, 
Miller's refined analysis of the half-period, 
second-harmonic effect observed in the original experiment, 
and in the subsequent ones by Morley and Miller \cite{morley}, 
showed that all data were consistent with an
effective, {\it observable} velocity lying in the range 7-10 km/s. 
For comparison, the Michelson-Morley experiment gave a value
$v_{\rm obs} \sim 8.8 $ km/s for the noon observations 
and a value $v_{\rm obs} \sim 8.0 $ km/s for the evening observations.

Due to the importance of the issue, we have decided to re-analyze the 
original data obtained by Michelson and Morley and
re-calculate the values of $v_{\rm obs}$ for their 
experiment. Our findings completely 
confirm Miller's indication of an observable velocity 
$v_{\rm obs}\sim 8.4$ km/s in their data. 

In addition assuming, as in the pre-relativistic physics, 
the existence of a preferred reference frame, but
using Lorentz transformations to connect with the Earth's reference frame, 
it turns out that this $v_{\rm obs}$
corresponds to a {\it real} Earth's velocity, 
in the plane of the interferometer, 
$v_{\rm earth} \sim 201 \pm 12$ km/s. 
This value, which is remarkably consistent
with Miller's kinematically  calculated value
$v_{\rm earth} \sim 203 \pm 8$ km/s
\cite{miller}, suggests that the magnitude of the fringe
shifts is determined by the typical velocity of the Solar System within our
galaxy (and not, for instance, by its velocity $v_{\rm earth} \sim 336$ km/s
with respect to the centroid of the Local Group). 

We emphasize that the use of Lorentz transformations is absolutely crucial. 
In fact, in this case, differently from the classical predictions, the fringe
shifts measured with an interferometer filled with a dielectric medium 
of refractive index ${\cal N}_{\rm medium}$ are proportional to the Fresnel's
drag coefficient $1- 1/{\cal N}^2_{\rm medium}$. For this reason, a large
`kinematical' velocity $\sim 200$ km/s is seen, in an in-air-operating optical
system, as a small `observable' velocity $\sim 8.4$ km/s. 
At the same time, without using Lorentz transformation, there was no hope
to understand why, for the same value of $v_{\rm earth}$, 
the effective $v_{\rm obs}$ had to be $\sim 3$ km/s 
for the Illingworth experiment (performed in an apparatus filled with helium) or
$\sim 1$ km/s for the Joos experiment (performed in an evacuated housing).

\vskip 10 pt

\section{The Michelson-Morley data}

We have analyzed 
the original data obtained by Michelson and Morley in {\it each} of
the six different sessions
of their experiment. No form of 
inter-session averaging has been attempted. 
As discovered by Miller, in fact, 
inter-session averaging of the raw data may be misleading.
For instance, in the Morley-Miller data \cite{morley}, 
the morning and evening observations each were indicating
 an effective velocity of
about 7.5 km/s (see Fig.11 of Ref.\cite{miller}). This indication
 was completely lost
with the wrong averaging procedure adopted in Ref.\cite{morley}. 
The same point of view has been advocated by Munera in his recent 
re-analysis of the classical experiments \cite{munera}. 

To obtain the fringe shifts of each session we have
followed the well defined procedure adopted in the classical experiments as
described in Miller's paper \cite{miller}. Namely, starting from the seventeen
entries, say $E(i)$, reported in the Michelson-Morley Table \cite{mm}, 
one was first correcting the data for the difference $E(1) -E(17)$ 
between the 1st entry and the 17th entry
obtained after a complete rotation of the apparatus.
Therefore, assuming the linearity of the
correction effect, one was adding 15/16 of the correction
to the 16th entry, 14/16 to the
15th entry and so on, thus obtaining a set of 16 corrected entries 
\BE
E_{\rm corr}(i)={{i-1}\over{16}} (E(1)-E(17)) + E(i)
\EE
Finally, the fringe shift is defined from the 
differences between each of the corrected entries $E_{\rm corr}(i)$ and 
their average value $\langle E_{\rm corr} \rangle$ as
\BE
          {{\Delta \lambda (i)}\over{\lambda}}= E_{\rm corr}(i) - \langle E_{\rm corr} \rangle
\EE
These final data for each session are shown in Table 1. 

Following the above procedure, 
the fringe shifts are given as a periodic function (with vanishing mean)
in the range $0 \leq \theta \leq 2\pi$ with 
$\theta={{i-1}\over{16}} 2\pi$. Therefore, they can be reproduced in a
Fourier expansion
\BE
\label{fourier}
      {{\Delta \lambda(\theta)}\over{\lambda}} =\sum_n~\bar{A}_n 
\cos(n\theta+ \phi_n)
\EE
In this way, one can extract
the amplitude $\bar{A}_2$ of 
the second-harmonic component which is the relevant one to determine the 
observable velocity. 
Following Miller's indications,  we have included terms up 
to $ n=5$, although 
the results for $\bar{A}_2$ are practically unchanged if one excludes from 
the fit the terms with $n=4$ and $n=5$. 
The typical fit to the data 
is illustrated in Fig.1  while we report
in Table 2 the values of $\bar{A}_2$ for each session.

The Fourier analysis allows to 
determine the azimuth of the ether-drift effect, 
from the phase $\phi_2$ of the second-harmonic component, and an observable velocity 
from the value of its amplitude. 
To this end, we have used the basic relation of the experiment
\BE
                 2 \bar{A}_2= 
{{2D}\over{\lambda}} ~{{v^2_{\rm obs} }\over{c^2}}
\EE
where $D$ is the length of each arm of the interferometer. 

Notice that, as emphasized by Shankland et al. (see page 178 of 
Ref.\cite{shankland}), it is the quantity 
$2 \bar{A}_2$, and not $\bar{A}_2$ itself, 
that should be compared with the maximal
displacement obtained for rotations of the apparatus through 
$90^o$ in its optical plane (see also Eqs.(\ref{fringe}) and (\ref{abar2})
below). Notice also that the quantity $2\bar{A}_2$ is denoted by $d$ in 
Miller's paper (see page 227 of Ref.\cite{miller}). 

 Therefore, 
for the Michelson-Morley apparatus where
 ${{D}\over{\lambda}}\sim 2\cdot 10^7$ \cite{mm}, 
 it becomes convenient to normalize the experimental values of
$\bar{A}_2$ to the classical prediction for an Earth's velocity of 30 
km/s 
\BE
\label{classical}
{{D}\over{\lambda}} ~{{(30 {\rm km/s})^2 }\over{c^2}}\sim 0.2
\EE
and we obtain
\BE
v_{\rm obs}  \sim 30 ~ \sqrt { 
{\bar{A}_2 }\over{ 0.2 } }~~{\rm km/s} 
\EE
Now, by inspection of Table 1, we find that
the average value of $\bar{A}_2$ from the noon sessions, 
$\bar{A}_2=0.017 \pm 0.003$, indicates a velocity
$v_{\rm obs}=8.7 \pm 0.8$ km/s
and the average value from the evening sessions, $\bar{A}_2=0.014 \pm 0.003$, 
indicates a velocity $v_{\rm obs}=8.0 \pm 0.8$ km/s. Since 
the two determinations
are well consistent with each other, we conclude that the 
Michelson-Morley experiment provides an ${\bar A}_2$ which is $\sim 1/13$ of the
classical expectation and an observable velocity
\BE
\label{vobs}
       v_{\rm obs} = 8.4 \pm 0.5~{\rm km/s}
\EE
in excellent agreement with Miller's estimate.

Notice that this value is also in excellent
agreement with the results obtained by Miller {\it himself} at Mt. Wilson. 
 Differently from the original Michelson-Morley experiment, 
Miller's data were taken over the 
entire day and in four epochs of the year. 
However, after the critical re-analysis of Shankland et al. \cite{shankland}, 
it turns out that
the average daily determinations of $\bar{A}_2$ for the four epochs
were statistically consistent (see page 170 of Ref.\cite{shankland}).
Therefore, one can take
the average of the four daily determinations, $\bar{A}_2=0.044 \pm 0.005$, 
and compare with the equivalent form of Eq.(\ref{classical})
for the Miller's interferometer
${{D}\over{\lambda}} ~{{(30 {\rm km/s})^2 }\over{c^2}}\sim 0.56$. Again, 
the observed $\bar{A}_2$ is just $\sim 1/13$ of 
the classical expectation for an Earth's velocity of 30 km/s and 
the effective 
$v_{\rm obs}$ is {\it exactly} the same as in Eq.(\ref{vobs}). 

\section{Miller's 1932 cosmic solution}

The problem with Miller's analysis was to reconcile such low observable
 values of the Earth's velocity 
with those obtained from the daily {\it variations} of the azimuth ( i.e.
$\phi_2$) and magnitude (i.e. $\bar{A}_2$) of the ether-drift effect. 
In this way,
on the base of the theory exposed by Nassau and Morse \cite{morse}, 
Miller could obtain two determinations of the apex of the Earth's motion 
for any given epoch of the year. These consist in two
pairs of values 
$(\alpha,\delta)$ where $\alpha$ denotes 
the right ascension and $\delta$ the declination. 
The pair of values obtained from the variation of the azimuth, 
say $(\alpha$-Az,$\delta$-Az), and that obtained
 from the variation of the magnitude,
say $(\alpha$-Mag,$\delta$-Mag), were found 
in good agreement with each
other (see Fig.23 of Ref.\cite{miller}). Therefore, it makes sense 
to average the two determinations in a single value, 
say 
$(\langle\alpha\rangle,\langle\delta\rangle)$, 
for each of the four epochs of the year 
of his observations. These
four average values lie, to a very good approximation, 
on the Earth's `aberration orbit', the centre of which 
is the apex of the cosmic component of the Earth's motion. Subtracting out from
each $(\langle\alpha\rangle,\langle\delta\rangle)$ the 
known effects of the Earth's orbital motion, 
Miller could finally restrict kinematically
the cosmic component of the Earth's velocity in the range 200-215
km/s (see page 233 of Ref.\cite{miller}) with the conclusion that 
"...a velocity $ v_{\rm earth} \sim 208 ~ {\rm km/s}$ 
for the cosmic component, gives the closest 
grouping of the four independently determined locations of the cosmic 
apex". The direction of the apex thus determined points toward the midst of
the Great Magellanic Cloud $20^o$ south of Canopus, the second brightest star
in the heavens. 

At the same time, due to the
particular magnitude and direction of the cosmic component, Miller's 
predictions for the total Earth's velocity 
in the plane of the interferometer had very 
similar values (see Table V of Ref.\cite{miller}), say
\BE
\label{vcosmic}
v_{\rm earth} \sim 203 \pm 8 ~ {\rm km/s}
\EE
Therefore, after Miller's observations, the situation with the ether-drift 
experiments could be summarized as follows (see page 236 of Ref.\cite{miller}).
On one hand, "the observed displacement of the interference fringes, for some
unexplained reason, corresponds  to only a fraction of the velocity of the
Earth in space". On the other hand, the theoretical solution of the Earth's 
cosmic motion involves only the relative values of the ether-drift effect
and "..does not require a knowledge of the cause of the reduction in the
apparent velocity nor of the amount of this reduction". A check of this is 
that, after inserting the final parameters of the cosmic component in the
Nassau-Morse expressions, "..the calculated curves fit the observations 
remarkably well, considering the nature of the experiment" (see Figs. 26 and 27
of Ref.\cite{miller}). 

In spite of this beautiful agreement, 
the unexplained large discrepancy between the typical values of $v_{\rm obs}$, 
as given in Eq.(\ref{vobs}), and the typical calculated
values of $v_{\rm earth}$, as
given in Eq.(\ref{vcosmic}), has been representing a very
serious objection to the consistency of Miller's analysis. 

\section{The role of Lorentz transformations}

It has been recently 
pointed out, however, by Cahill and Kitto \cite{cahill} that an effective
reduction of the Earth's velocity, from a large `kinematical' value 
$v_{\rm earth}={\cal O}(10^2)$ km/s down to a small `observable' value
$v_{\rm obs}={\cal O}(1)$ km/s, can be understood by
taking into
account the effects of the Lorentz contraction and of the
refractive index ${\cal N}_{\rm medium}$ of the
dielectric medium used in the interferometer. 

In this way, the
observations become consistent \cite{cahill} with values of the
Earth's velocity that are comparable to 
$v_{\rm earth} \sim 365$
km/s as extracted by fitting the COBE data for the cosmic
background radiation \cite{cobe}. The point is that the fringe shifts are
proportional to $ {{v^2_{\rm earth} }\over{c^2}}
 (1- {{1}\over{ {\cal N}^2_{\rm medium} }})$ rather than to
                  ${{v^2_{\rm earth} }\over{c^2}}$ itself. For the air, 
where ${\cal N}_{\rm air}\sim 1.00029$, assuming a value
$v_{\rm earth} \sim 365$ km/s, one would expect fringe shifts 
governed by an effective velocity 
$v_{\rm obs}\sim 8.8 $ km/s consistently with our value
Eq.(\ref{vobs}).

This would also explain why the experiments
of Illingworth \cite{illing} (performed in an apparatus filled with helium 
where ${\cal N}_{\rm helium}\sim 1.000036$) 
and Joos \cite{joos} (performed in the vacuum where
${\cal N}_{\rm vacuum}\sim 1.00000..$)
were showing smaller fringe shifts and, therefore, 
lower effective velocities. 

In Ref.\cite{consoli} the argument has been completely reformulated by
 using Lorentz transformations (see also Ref.\cite{pagano}). 
As a matter of fact, in 
this case there is a non-trivial difference of a factor $\sqrt{3}$. 
When properly taken into account, the Earth's velocity extracted from
the absolute magnitude of the fringe shifts is {\it not} 
$v_{\rm earth} \sim 365$ km/s but $v_{\rm earth} \sim 201$ km/s thus
making Miller's prediction Eq.(\ref{vcosmic}) completely consistent with 
Eq.(\ref{vobs}). For the  convenience of the reader, we shall report in the
following the essential steps. 

The key point
is that Lorentz transformations preserve the value of the speed of light 
{\it in the vacuum} $c=2.9979..10^{10}$ cm/s
but do not preserve its value 
\BE
\label{u}
u\equiv  
{{c}\over{\cal{N}_{\rm medium} }}
\EE
in a medium. In this case, due to a
refractive index ${\cal N}_{\rm medium} > 1$, one has to account for the
effects of a non-vanishing Fresnel's drag coefficient
\BE
 k_{\rm medium}=
1- {{1}\over{ {\cal N}^2_{\rm medium} }} \ll 1
\EE
Therefore, if light would be seen to propagate isotropically 
with velocity Eq.(\ref{u}) in one 
(`preferred') reference frame $\Sigma$, it  
will not be seen to propagate isotropically 
 in any other frame $S'$ that is in relative motion with respect to
$\Sigma$. 

Now, this is precisely the basic issue: determining experimentally, and
to a high degree of accuracy, whether light propagates
isotropically  for an observer $S'$ placed on the
Earth. For instance within the air, where the relevant value is
${\cal N}_{\rm air}=1.00029..$, the isotropical value
${{c}\over{\cal{N}_{\rm air} }}$ is usually determined directly by
measuring the two-way speed of light along various directions. In this way, 
isotropy can be established at the level $\sim 10^{-7}$. 
If we require, however, 
a higher level of accuracy, say $10^{-9}$, the only way 
to test isotropy is to 
perform a Michelson-Morley type of experiment and look for 
fringe shifts upon rotation of the interferometer. 

Therefore, if one finds experimentally fringe shifts (and thus 
some non-zero anisotropy) it becomes natural 
to explore the possibility that this effect is due 
to the Earth's motion with respect to a preferred frame 
$\Sigma\neq S'$.
Assuming this scenario, the degree of anisotropy for $S'$ 
can easily be determined by using Lorentz
transformations. By denoting
${\bf{v}}$ the velocity of
$S'$ with respect to $\Sigma$ one finds 
($\gamma= 1/\sqrt{ 1- {{ {\bf{v}}^2}\over{c^2}} }$) 
\BE
\label{uprime}
  {\bf{u}}'= {{  {\bf{u}} - \gamma {\bf{v}} + {\bf{v}}
(\gamma -1) {{ {\bf{v}}\cdot {\bf{u}} }\over{v^2}} }\over{ 
\gamma (1- {{ {\bf{v}}\cdot {\bf{u}} }\over{c^2}} ) }}
\EE
where $v=|{\bf{v}}|$. By keeping terms up 
to second order in $v/u$, denoting by
$\theta$ the angle between ${\bf{v}}$ and ${\bf{u}}$
 and defining $u'(\theta)= |{\bf{u'}}|$, 
we obtain
\BE
  {{ u'(\theta) }\over{u}}= 1- \alpha {{v}\over{u}} -\beta {{v^2}\over{u^2}}
\EE
where 
\BE
   \alpha = (1-  {{1}\over{ {\cal N}^2_{\rm medium} }} ) \cos \theta + 
{\cal O} ( ({\cal N}^2_{\rm medium}-1)^2 )
\EE
\BE
\beta = (1- {{1}\over{ {\cal N}^2_{\rm medium} }} ) P_2(\cos \theta) +
{\cal O} ( ({\cal N}^2_{\rm medium}-1)^2 )
\EE
with $P_2(\cos \theta) = {{1}\over{2}} (3 \cos^2\theta -1)$.

Finally, the two-way speed of light is 
\BE
\label{twoway}
{{\bar{u}'(\theta)}\over{u}}= {{1}\over{u}}~ {{ 2  u'(\theta) u'(\pi + \theta) }\over{ 
u'(\theta) + u'(\pi + \theta) }}= 1- {{v^2}\over{c^2}} ( A + B \sin^2\theta) 
\EE
where 
\BE 
\label{ath}
   A= {\cal N}^2_{\rm medium} -1 + {\cal O} ( ({\cal N}^2_{\rm medium}-1)^2 )
\EE
and 
\BE
\label{BTH}
     B= -{{3}\over{2}} 
({\cal N}^2_{\rm medium} -1 )
+ {\cal O} ( ({\cal N}^2_{\rm medium}-1)^2 )
\EE
To address the theory of the Michelson-Morley
interferometer we shall consider
two light beams, say 1 and 2, that for simplicity are chosen 
perpendicular in $\Sigma$
where they propagate along the $x$ and $y$ axis with velocities
$u_x(1)=u_y(2)=
u= {{c}\over{ {\cal N}_{\rm medium} }}$.
Let us also assume that the velocity $v$ of  $S'$ is along the $x$ axis.

Let us now define $L'_P$ and $L'_Q$ to be the lengths of two
optical paths, say P and Q, as
measured in the $S'$ frame. For instance, they can represent the
lengths
of the arms of an interferometer which is at rest in the $S'$ frame.
In the first experimental set-up, the arm
of length $L'_P$ is taken along the direction of motion associated with the beam
1 while the arm of length $L'_Q$ lies along the direction of the beam 2.

In this way, the interference pattern, between the light beam coming out
of the optical path P and that coming out of the optical path Q, 
 can easily be obtained from the relevant delay time. 
 By using the
equivalent form of the Robertson-Mansouri-Sexl parametrization
\cite{robertson,mansouri} for the two-way speed of light 
defined above in Eq.(\ref{twoway}), this is given by
\BE
   \Delta T'(0) =
{{2L'_P}\over{\bar{u'}(0)}}- {{2 L'_Q}\over{\bar {u'}(\pi/2)}} 
\EE
On the other hand, if the beam 2 were to propagate along the optical path
P and
the beam 1 along Q, one would obtain a different
delay time, namely
\BE
   {(\Delta T')}_{\rm rot}=
{{2L'_P}\over{\bar{u'}(\pi/2)}}- {{2 L'_Q}\over{\bar {u'}(0)}} 
\EE
Therefore, by rotating the apparatus and
using Eqs.(\ref{ath}) and (\ref{BTH}), 
one obtains fringe shifts proportional to
\BE 
\label{deltat2} \Delta T'(0)- (\Delta T')_{\rm
rot} \sim (-2B) {{(L'_P+ L'_Q)}\over{u}}  
{{v^2}\over{u^2}} 
\EE 
or
\BE 
\label{deltat} 
\Delta T'(0)- (\Delta T')_{\rm
rot} \sim {{3(L'_P+ L'_Q)}\over{u}}  k_{\rm medium}
{{v^2}\over{u^2}} 
\EE 
(neglecting ${\cal O}(\kappa^2_{\rm medium})$ terms). 
This coincides with the pre-relativistic expression provided one replaces
$v$ with an effective observable velocity
\BE
\label{vobs0}
           v_{\rm obs}= v
\sqrt { k_{\rm medium} }
 \sqrt{3}
\EE
Finally, 
for the Michelson-Morley experiment, where $L'_P=L'_Q=D$, and for an ether wind
along the $x$ axis, 
the prediction for the fringe shifts at a given angle 
$\theta$ has the particularly simple form
\BE 
\label{fringe}
{{\Delta \lambda (\theta)}\over{\lambda}}=
  {{u \Delta T'(\theta) }\over{\lambda}} = {{u}\over{\lambda}} (
{{2D }\over{\bar{u'}(\theta)}}- {{2 D }\over{\bar {u'}(\pi/2+\theta)}})=
 {{2 D }\over{\lambda}} {{v^2}\over{c^2}} (-B) \cos (2\theta) 
\EE
that corresponds to a pure second-harmonic 
effect. At the same time, it becomes clear the remark by 
Shankland et al. (see page 178 of Ref.\cite{shankland})
that its amplitude 
\BE
\label{abar2}
              \bar{A}_2 \equiv  
{{2 D }\over{\lambda}} {{v^2}\over{c^2}} (-B) = 
{{D }\over{\lambda}} {{v^2_{\rm obs} }\over{c^2}}  
\EE
is just one-half of the corresponding quantity entering Eq.(\ref{deltat2}). 

\section{Interpretation of the Michelson-Morley observations}

Now, if upon operation of the interferometer
there are fringe shifts  and if their magnitude,
 observed with different dielectric media and within the
experimental errors, points
consistently to a unique value of the Earth's velocity, there is
experimental evidence for
the existence of a preferred frame $\Sigma \neq S'$. In practice, to
 ${\cal O}({{v^2_{\rm earth} }\over{c^2}} )$, this can be decided by
re-analyzing \cite{cahill}
the experiments in terms of the effective parameter
 $\epsilon = {{v^2_{\rm earth} }\over{u^2}} k_{\rm medium}$. The
conclusion of Cahill and Kitto \cite{cahill} 
is that the classical experiments are consistent with the value
 $v_{\rm earth}\sim 365$ km/s obtained from the COBE data. 

However, in
our expression Eq.(\ref{vobs0}) determining the fringe shifts there is a difference
of a factor $\sqrt{3}$ with respect to their result
$v_{\rm obs}=v \sqrt { k_{\rm medium} }$. Therefore, using
Eqs.(\ref{vobs0}) and (\ref{vobs}), for ${\cal N}_{\rm air} \sim 1.00029$, 
 the relevant Earth's velocity (in the plane of the interferometer)
 is {\it not} $v_{\rm earth}\sim 365$ km/s but rather
\BE
\label{vearth}
                  v_{\rm earth} \sim 201 \pm 12 ~{\rm km/s}
\EE
This value, obtained from the magnitude of the fringe shifts, is
in excellent agreement with the value Eq.(\ref{vcosmic}) calculated 
kinematically by Miller. 
Therefore, from this excellent agreement, we deduce that
the observed fringe shifts are determined by the typical
velocity of the Solar System within our galaxy and not, for instance, 
by its velocity relatively to the
centroid of the Local Group. 
In the latter case, one would get higher values such as
$v_{\rm earth} \sim 336$ km/sec, see Ref.\cite{V5}. 

Notice that such ambiguity, say  
$v_{\rm earth}\sim 200, 300, 365,...$ km/s, on the actual value of the
Earth's velocity determining
the fringe shifts, can only be resolved experimentally in view of the
many theoretical uncertainties in the operative
definition of the preferred frame
where light propagates isotropically. At this stage, we believe, one should 
just concentrate on the internal consistency of the various frameworks. 
In this sense, the analysis presented in this paper shows
that internal consistency is extremely high in Miller's 1932 solution. 

We are aware that our conclusion goes against the widely spread belief 
that Miller's results were only due to statistical fluctuation and/or 
local temperature conditions (see the Abstract of Ref.\cite{shankland}). 
However, within the paper the same authors of Ref.\cite{shankland}
say that "...there can be 
little doubt that statistical fluctuations alone  cannot account for the
periodic fringe shifts observed by Miller" 
(see page 171 of Ref.\cite{shankland}).  In fact, although "...there is 
obviously considerable scatter in the data at each azimuth position,...the
average values...show a marked second harmonic effect"
(see page 171 of Ref.\cite{shankland}). In any case, interpreting the observed
effects on the base of the local temperature conditions cannot be the 
whole story since "...we must admit that a direct and general 
quantitative correlation between amplitude and phase of the observed 
second harmonic on the one hand and the thermal conditions in the observation
hut on the other hand could not be established" 
(see page 175 of Ref.\cite{shankland}). This rather
unsatisfactory explanation of the observed effects should be compared 
with the previously mentioned
excellent agreement that was instead obtained by Miller
once the final parameters for the Earth's velocity were plugged in the 
theoretical predictions (see Figs.26 and 27 of Ref.\cite{miller}).

The most surprising thing, however, is that Shankland et al. did not realize
that Miller's average value
$\bar{A}_2=0.044 \pm 0.005$, obtained after their own critical re-analysis
of his observations, gives precisely the same $v_{\rm obs}$ 
Eq.(\ref{vobs}) obtained from the Miller's re-analysis of the Michelson-Morley 
experiment. Conceivably, 
their emphasis on the role of the temperature effects
in Miller's data would have been turned down
whenever they had realized the perfect identity
between two determinations obtained 
in completely different experimental
conditions. 

\section{Comparison with other classical experiments}

On the other hand, 
additional information on the validity of Miller's results
can also be obtained by other means, for instance 
comparing with the experiment performed by Michelson, 
Pease and Pearson \cite{mpp}. These other authors in 1929, 
using their own interferometer, again at Mt. Wilson, declared that 
their "precautions taken to eliminate effects of 
temperature and flexure disturbances were effective". Therefore, their statement that the
fringe shift, 
as derived from "..the displacements observed at maximum and minimum at 
sidereal times..", was definitely smaller than "...one-fifteenth of
that expected on the 
supposition of an effect due to a motion of the Solar System of three 
hundred kilometres per second", can be taken as an indirect 
confirmation of Miller's results. Indeed,
although the "one-fifteenth" was actually 
a "one-fiftieth" (see pag.240 of Ref.\cite{miller}), 
their fringe shifts were certainly non negligible. This is easily understood 
since, for an in-air-operating interferometer, 
the fringe shift $(\Delta\lambda)_{\rm class}(300)$, expected on the base of 
classical physics
for an Earth's velocity of 300 km/s, is about 500 times
bigger than the corresponding relativistic one
\BE
(\Delta\lambda)_{\rm rel}(300)\equiv 3 k_{\rm air}
~ (\Delta\lambda)_{\rm class}(300)
\EE
computed using Lorentz transformations 
(compare with Eqs.(\ref{deltat}) and (\ref{vobs0}) for 
$k_{\rm air}\sim {\cal N}^2_{\rm air} -1 \sim 0.00058$). 
 Therefore, the Michelson-Pease-Pearson upper bound
\BE
(\Delta\lambda)_{\rm obs}< 0.02~
 (\Delta\lambda)_{\rm class} (300)
\EE
is actually equivalent to
\BE
(\Delta\lambda)_{\rm obs}< 24 ~
 (\Delta\lambda)_{\rm rel} (208)
\EE
As such, it poses no strong restrictions and is entirely 
consistent with those typical low effective velocities detected
by Miller in his observations of 1925-1926. 

A similar agreement is obtained when comparing with the Illingworth's data
\cite{illing} as recently 
re-analyzed by Munera \cite{munera}. In this case, using Eq.(\ref{vobs0}), 
from
the observable velocity $v_{\rm obs}=3.13 \pm 1.04$ km/s \cite{munera} 
and the value
${\cal N}_{\rm helium} \sim 1.000036$, 
 one deduces $v_{\rm earth}=213 \pm 71$ km/s, in very good agreement with 
Miller's calculated value Eq.(\ref{vcosmic}). 

The same conclusion applies to the Joos experiment 
\cite{joos}. His interferometer was placed in an evacuated housing and
he declared that the velocity of any ether wind 
had to be smaller than 1.5 km/s.
Although we don't know the exact 
value of ${\cal N}_{\rm vacuum}$ for the Joos experiment, 
it is clear that this is the type of upper bound
expected in this case. As an example, for $v_{\rm earth}\sim 208$ km/s, one
obtains $v_{\rm obs}\sim 1.5$ km/s for 
${\cal N}_{\rm vacuum}-1= 9\cdot 10^{-6}$ and 
$v_{\rm obs}\sim 0.5$ km/s for 
${\cal N}_{\rm vacuum}- 1=1\cdot 10^{-6}$. 
In this sense, the effect of using Lorentz 
transformations is most 
dramatic for the Joos experiment when comparing with
the classical expectation for an Earth's velocity
of 30 km/s. Although the relevant Earth's velocity 
is $\sim 208$ km/s, 
the fringe shifts, rather than being $\sim 50$ times {\it bigger} than the 
classical prediction, are  $\sim 1000$ times {\it smaller}. 

\section{Comparison with present-day experiments}

Lets us now briefly address a comparison with present-day, 
 `high vacuum' Michelson-Morley experiments of the type first performed by 
Brillet and Hall \cite{brillet} and more recently by 
M\"uller et al. \cite{muller}. In a perfect vacuum, by definition 
${\cal N}_{\rm vacuum}=1$ so that $v_{\rm obs}=0$ and no anisotropy can 
be detected. However, one can explore 
\cite{pagano,modern} the possibility that, even in this case,
 a very small anisotropy might be due to a refractive index 
${\cal N}_{\rm vacuum}$ that differs from unity by an infinitesimal
amount. In this case, the natural candidate to explain a value
${\cal N}_{\rm vacuum} \neq 1$
is gravity. In fact, by using
the Equivalence Principle, any freely falling frame $S'$ will locally 
measure the same speed of light as in an inertial frame in the absence of
any gravitational effects. However, if $S'$ carries on board an heavy 
object this is no longer true. For an observer placed on the Earth, 
this amounts to insert
the Earth's gravitational potential in the  weak-field isotropic
approximation to the line element of
 General Relativity \cite{weinberg}
\BE
ds^2= (1+ 2\varphi) dt^2 - (1-2\varphi)(dx^2 +dy^2 +dz^2)
\EE
so that one obtains a refractive index for
light propagation 
\BE
\label{nphi}
            {\cal N}_{\rm vacuum}\sim  1- 2\varphi
\EE
This represents the `vacuum analogue' of 
${\cal N}_{\rm air}$, ${\cal N}_{\rm helium}$,...so that from
\BE 
     \varphi =- {{G_N M_{\rm earth}}\over{c^2 R_{\rm earth} }} \sim
-0.7\cdot 10^{-9}
\EE
and using Eq.(\ref{BTH}) one predicts
\BE
\label{theor}
                 B_{\rm vacuum} \sim -4.2 \cdot 10^{-9}
\EE
For $v_{\rm earth} \sim 208$ km/s, 
this implies an observable anisotropy of the two-way speed of light 
in the vacuum Eq.(\ref{twoway}) 
\BE
        {{ \Delta \bar{c}_\theta }\over{c}} \sim 
|B_{\rm vacuum}| {{v^2_{\rm earth} }\over{c^2}} \sim 2\cdot 10^{-15}
\EE
in good agreement with the experimental value
        ${{ \Delta \bar{c}_\theta }\over{c}}= (2.6 \pm 1.7) \cdot 10^{-15}$ 
determined by M\"uller et al.\cite{muller}. 

\section{Summary and conclusions}

In this paper we have presented our re-analysis of the original data obtained
by Michelson and Morley \cite{mm}. 
Contrary to the generally accepted ideas, 
 but in agreement with the point of view 
expressed by Hicks in 1902, Miller in 1933 and Munera in 1998,
the results of that experiment
cannot be considered null. The observed
fringe shifts, although smaller than the classical prediction corresponding
to the orbital motion of the Earth, point to an effective 
observable velocity $v_{\rm obs}=8.4 \pm 0.5$ km/s. As emphasized at the end
of Sect.2 and at the end of Sect.5, 
this value is exactly the
same average value obtained by Miller in his observations
at Mt.Wilson (after the critical 
re-analysis of Shankland et al.\cite{shankland}).

Therefore, it becomes natural to explore the existence
of a preferred reference frame and use Lorentz
transformations to extract the real kinematical velocity corresponding to this
$v_{\rm obs}$. In this case, we find a value
(in the plane of the interferometer)
$v_{\rm earth}=201 \pm 12 $ km/s. This value 
is in excellent agreement with the
value that was kinematically calculated
by Miller on the base of the 
observed variations of the ether-drift effect in different epochs of the year.
As a consequence, we conclude that 
the magnitude of the fringe shifts is determined by the
typical velocity of the Solar System within our galaxy. This conclusion is
also consistent with the most recent data for
the anisotropy of the two-way speed of light in the vacuum.

It is somewhat surprising that
the crucial role of Lorentz transformations has been overlooked for so long 
time in the analysis of the Michelson-Morley experiment. 
This is, probably, due to the wrong impression that there should be very little
difference from the classical predictions since
 the velocities in the game (30, 200, 300,..) km/s are so 
much smaller than the speed of
light. The point is, however, that Lorentz transformations preserve 
exactly the value of the speed of light
in the vacuum. Therefore, in a medium
any anisotropy has to be proportional 
to the Fresnel drag coefficient $k_{\rm medium} \ll 1$. For an in-air-operating
optical interferometer, this  
effect is responsible for the huge reduction of the Earth's velocity from
its relevant
kinematical value $v_{\rm earth}\sim 200 $ km/s down to its observable value 
$v_{\rm obs}\sim 8.4$ km/s governing the magnitude of the fringe shifts in 
the Michelson-Morley and Miller's experiments.
At the same time, without using Lorentz transformation, there was no hope
to understand why, 
for the same value of $v_{\rm earth}$, 
the effective $v_{\rm obs}$ had to be $\sim 3$ km/s 
for the Illingworth experiment  or
$\sim 1$ km/s
for the Joos experiment.

This set of puzzling experimental results can now be described within a single
consistent framework. 
Therefore, we conclude that an alternative interpretation of the
Michelson-Morley type of experiments is now emerging. Contrary to the 
point of view that was generally accepted in the past century, 
they provide experimental
evidence for the existence of a preferred reference frame.

\vfill
\eject
\textwidth 24.0cm
\oddsidemargin  -0.20cm
\evensidemargin 0.0cm
\topmargin -0.8cm
\textheight 18.0cm  
\begin{table}
\centering{\begin{tabular}{c||c|c|c|c|c|c}  
 i      &  July 8 (n.) & July 9 (n.) & July 11 (n.)     
       &  July 8 (e.) & July 9 (e.) & July 12 (e.)     \\
1      & -0.001 & +0.018 &+0.015& -0.016& +0.007& +0.034 \\
2      & +0.024 & -0.004 &-0.035& +0.008& -0.015& +0.042 \\    
3      & +0.053 & -0.004 &-0.039& -0.010& +0.006& +0.045 \\   
4      & +0.015 & -0.003 &-0.067& +0.070& +0.004& +0.025 \\   
5      & -0.036 & -0.031 &-0.043& +0.041& +0.027& -0.004 \\  
6      & -0.007 & -0.020 &-0.015& +0.055& +0.015& -0.014 \\ 
7      & +0.024 & -0.025 &-0.001& +0.057& -0.022& +0.005 \\   
8      & +0.026 & -0.021 &+0.027& +0.029& -0.036& -0.013 \\    
9      & -0.021 & -0.049 &+0.001& -0.005& -0.033& -0.030 \\   
10     & -0.022 & -0.032 &-0.011& +0.023& +0.001& -0.066 \\    
11     & -0.031 & +0.001 &-0.005& +0.005& -0.008& -0.093 \\ 
12     & -0.005 & +0.012 &+0.011& -0.030& -0.014& -0.059 \\     
13     & -0.024 & +0.041 &+0.047& -0.034& -0.007& -0.040 \\       
14     & -0.017 & +0.042 &+0.053& -0.052& +0.015& +0.038 \\      
15     & -0.002 & +0.070 &+0.037& -0.084& +0.026& +0.057 \\      
16     & +0.022 & -0.005 &+0.005& -0.062& +0.024& +0.041 \\      
17     & -0.001 & +0.018 &+0.015& -0.016& +0.007& +0.034 \\     
\end{tabular}
\vskip 60 pt
\caption{\rm
We report the fringe shifts ${{\Delta \lambda(i)}\over{\lambda}}$
for all noon (n.) and evening (e.) sessions of the Michelson-Morley experiment.}
\label{tab:1}}
\end{table}
\vfill
\eject
\textwidth 15.5cm
\oddsidemargin 0.75cm
\evensidemargin 0.75cm
\topmargin -0.8cm
\textheight 21.5 cm  
\begin{table}
\centering{\begin{tabular}{c|c}  
SESSION       &       $\bar{A}_2$   \\
July 8  (noon) & $0.010 \pm 0.005$  \\ 
July 9  (noon) & $0.015 \pm 0.005$   \\
July 11 (noon) & $0.025 \pm 0.005$    \\
July 8  (evening) & $0.014 \pm 0.005$  \\
July 9  (evening) &$0.011 \pm 0.005$   \\
July 12 (evening) & $0.018 \pm 0.005$  \\ 
\end{tabular}
\vskip 20 pt
\caption{\rm
We report the amplitude of the second-harmonic component
$\bar{A}_2$ obtained from the fit Eq.(\ref{fourier}) 
to the various samples of data.}
\label{tab:2}}
\end{table}

\begin{figure}[ht]
\psfig{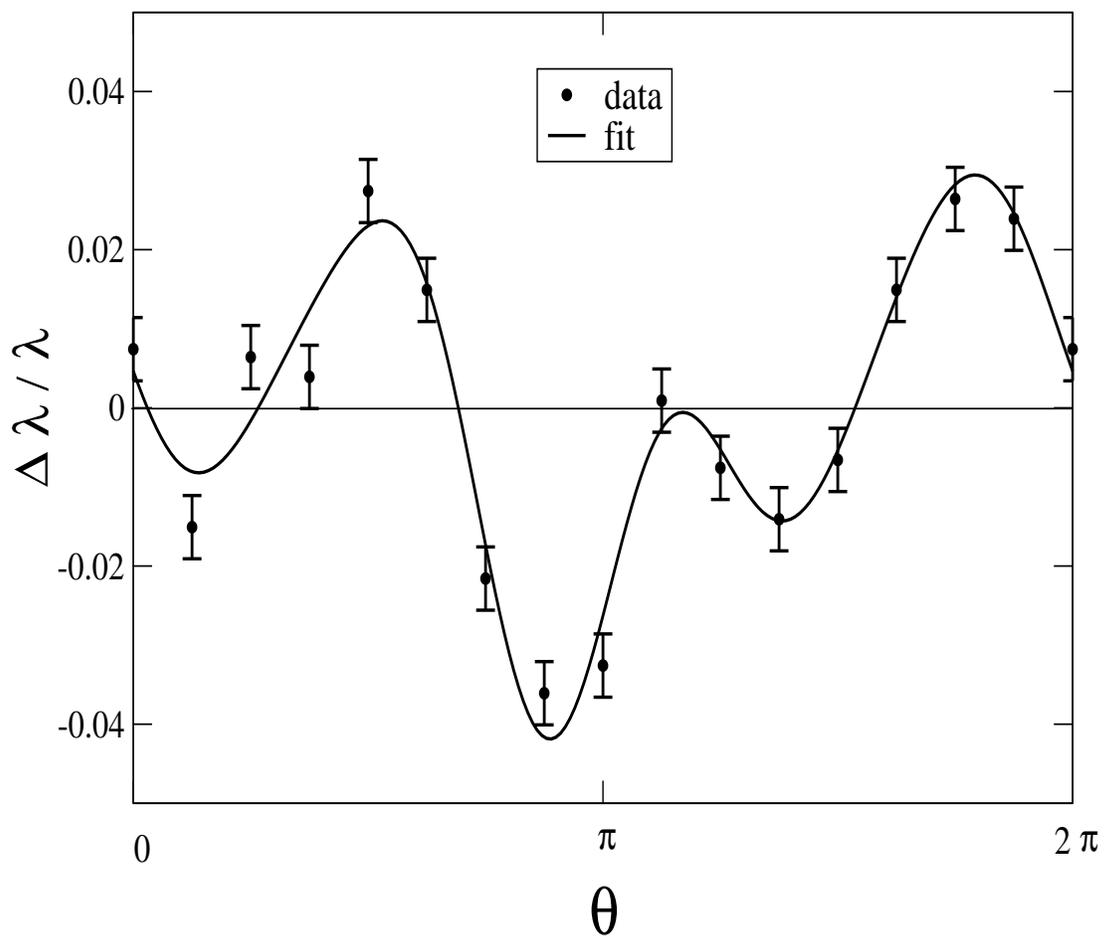}
\caption{
We show a typical fit Eq. (3) to the Michelson-Morley data shown in Table 1. 
}
\end{figure}

\vskip 60 pt

\begin{thebibliography} {99}
 \bibitem{mm}
A. A. Michelson and E. W. Morley, Am. J. Sci. {\bf 34}, 333 (1887).
 \bibitem{hicks}
W. M. Hicks, Phil. Mag.{\bf 3}, 9 (1902).
 \bibitem{miller}
D. C. Miller, Rev. Mod. Phys. {\bf 5}, 203 (1933).
 \bibitem{morley}
E. W. Morley and D. C. Miller, Phil. Mag. {\bf 9}, 669 (1905).
 \bibitem{munera}
H. A. Munera, APEIRON {\bf 5}, 37 (1998). 
 \bibitem{shankland}
R. S. Shankland et al., Rev. Mod. Phys. {\bf 27}, 167 (1955). 
 \bibitem{morse}
J. J. Nassau and P. M. Morse, ApJ. {\bf 65}, 73 (1927).
 \bibitem{cahill}
R. T. Cahill and K. Kitto, arXiV:physics/0205070.
 \bibitem{cobe}
G. F. Smoot et al., ApJL. {\bf 371}, L1 (1991).
 \bibitem{illing}
K. K. Illingworth, Phys. Rev. {\bf 30}, 692 (1927). 
 \bibitem{joos}
G. Joos, Ann. d. Physik {\bf 7}, 385 (1930).
 \bibitem{consoli}
M. Consoli, arXiv:physics/0310053, submitted to Physics Letters {\bf A}. 
 \bibitem{pagano}
M. Consoli, A. Pagano and L. Pappalardo, Phys. Lett. {\bf A318}, 292 (2003).
 \bibitem{robertson}
H. P. Robertson, Rev. Mod. Phys. {\bf 21}, 378 (1949).
 \bibitem{mansouri}
 R. M. Mansouri and R. U. Sexl, Gen. Rel. Grav. {\bf 8}, 497 (1977).
 \bibitem{V5}
G. De Vaucouleurs  and W. L. Peters, ApJ. {\bf 248}, 395 (1981).
 \bibitem{mpp}
A. A. Michelson, F. G. Pease and F. Pearson, Nature {\bf 123}, 88 (1929). 
 \bibitem{brillet}
A. Brillet and J. L. Hall, Phys. Rev. Lett. {\bf 42}, 549 (1979).
 \bibitem{muller}
H. M\"uller, S. Herrmann, C. Braxmaier, S. Schiller and A. Peters,
arXiv:physics/0305117.
 \bibitem{modern}
M. Consoli, arXiv:gr-qc/0306105.
 \bibitem{weinberg}
S. Weinberg, {\it Gravitation and Cosmology}, John Wiley and Sons, Inc., 1972,
pag. 181.

\end{thebibliography}
\end{document}